%% file: iclr2021_conference.tex
\UseRawInputEncoding
\documentclass{article} 

\usepackage{iclr2021_conference,times}
\input{math_commands.tex}

\usepackage{xcolor}
\usepackage{enumitem}
\usepackage{bbm}
\usepackage{natbib}
\usepackage{amsfonts}
\usepackage{amssymb}
\usepackage{amsmath}
\usepackage{graphicx}
\usepackage{dcolumn}
\usepackage{bm}
\usepackage{wrapfig}
\usepackage{subcaption}
\usepackage{microtype}
\usepackage[normalem]{ulem}
\usepackage[ruled, vlined,inoutnumbered,rightnl]{algorithm2e}
\SetAlFnt{\normalsize}
\SetAlCapFnt{\normalsize}
\SetAlCapNameFnt{\normalsize}

\usepackage[pdfauthor={Sam Foreman},%
   pdftitle={Deep Learning Hamiltonian Monte Carlo},%
   unicode,%
   linktocpage=true,%
   hyperfootnotes=true,
   pdftex,
]{hyperref}

\usepackage{url}

\definecolor{caribbeangreen}{rgb}{0.0, 0.8, 0.6}
\definecolor{pinkred}{HTML}{FF2052}
\definecolor{newred}{HTML}{f92672}
\definecolor{newblue}{HTML}{007fff}
\definecolor{newdarkblue}{HTML}{0282CA}
\definecolor{newpurple}{HTML}{CC4FFF}
\definecolor{newgreen}{HTML}{51C569}
\definecolor{newlightblue}{HTML}{71C8A7}
\definecolor{linkblue}{HTML}{0282CA}
\definecolor{linkblue1}{HTML}{2196F3}
\definecolor{awesome}{HTML}{FF2052}
\definecolor{darkviolet}{rgb}{0.58, 0.0, 0.83}
\hypersetup{
  colorlinks=true,
  urlcolor=newblue,
  linkcolor=awesome,
  citecolor=newpurple,
}

\definecolor{blue1}{HTML}{448aff}
\definecolor{pink1}{HTML}{FA5477}

\newcommand{\mbart}{\textcolor{awesome}{\bar{m}^{k}}}
\newcommand{\mt}{\textcolor{newblue}{m^{k}}}

\title{Deep Learning Hamiltonian Monte Carlo}

\author{Sam Foreman\(^{1}\), Xiao-Yong Jin\(^{1,2}\) \& James C. Osborn\(^{1,2}\)
\thanks{
   \texttt{\href{mailto:foremans@anl.gov}{foremans@anl.gov}},
   \texttt{\href{mailto:xjin@anl.gov}{xjin@anl.gov}},
   \texttt{\href{mailto:osborn@alcf.anl.gov}{osborn@alcf.anl.gov}}
}\\
   \(^{1}\)Leadership Computing Facility, \(^{2}\)Computational Science Division\\
   Argonne National Laboratory\\
}

\iclrfinalcopy

\begin{document}

\maketitle
\begin{abstract}
   We generalize the Hamiltonian Monte Carlo algorithm with a stack of neural network layers,
   and evaluate its ability to sample from
   different topologies in a two-dimensional lattice gauge theory.
   We demonstrate that our model is able to successfully mix between modes of different topologies,
   significantly reducing the computational cost required to generate independent gauge field configurations.
   Our implementation is available at
   \texttt{\href{https://github.com/saforem2/lh2mc-qcd}{https://github.com/saforem2/l2hmc-qcd}}.
\end{abstract}

\section{\label{sec:introduction}Introduction}
In \citet{levy2017}, the authors propose the Learning to Hamiltonian Monte Carlo (L2HMC) algorithm, and demonstrate
its ability to outperform traditional Hamiltonian Monte Carlo (HMC) on a variety of 
target
distributions.
They show that the trained L2HMC sampler is capable of mixing between modes by exploring regions of phase space which are 
inaccessible with traditional HMC.\@
In this paper, we propose a generalized sampler using a deep neural network that is
self-trained to propose new Markov Chain states,
which is made exact with the help of Metropolis-Hastings \citep{zbMATH03349185} algorithm.
We target our Deep Learning Hamiltonian Monte Carlo (DLHMC) algorithm to simulating lattice gauge theories,
where state-of-the-art simulations have billions of degrees of freedom and run on supercomputers with thousands of nodes,
yet suffer an exponential slowdown when approaching continuum physics \citep{schaefer2009investigating,cossu2018testing}.

\section{\label{sec:maincontributions}Main Contributions}
We propose the DLHMC algorithm as a generalized HMC algorithm, using a stack of distinct neural network layers replacing
consecutive leapfrog steps.
Each distinct neural network layer, carrying a discrete index \(k = 0, 1, \ldots, N_{\mathrm{LF}}-1\) for
$N_{\mathrm{LF}}$ leapfrog layers, performs the augmented leapfrog equations (\Eqref{eq:newmomentumupdate} and
\Eqref{eq:newpositionupdate}), which are parametrized by different neural networks.

We apply the proposed method to a \(2\)-dimensional \(U(1)\) lattice gauge theory defined on a square lattice with
periodic boundary conditions.
We specifically designed a loss function for our physical system, encouraging the tunneling of topological properties.
We see a significant reduction in the computational cost using DLHMC, as measured by the integrated autocorrelation time
of the topological charge.
We compare our results to traditional HMC across a variety of trajectory lengths and coupling constants, and show that
our trained models consistently outperform traditional HMC (see \Figref{fig:autocorrvsbeta}).
Models trained on smaller coupling constants, when applied to a larger coupling constant, require only minimal
retraining.
We find that the efficiency in tunneling topological sectors of our trained DLHMC sampler
is explained by the behavior of the physical system as it passes through the deep neural network,
see \Figref{fig:plaqsf} and \Figref{fig:hwf}.
\section{\label{sec:relatedwork}Related Work}
Recently, there has been a growing interest in applying generative machine learning techniques to build smarter, more
efficient scientific simulations.
Following the development of the RealNVP \citep{dinhRealNVP} architecture, there have been many proposed techniques that
aim to take advantage of its invertible structure.
In particular, simulations in lattice gauge theory and lattice QCD stand to benefit tremendously from this new approach,
as evidenced by the rapidly-growing body of work in this direction ranging from network architectures
\citep{dinhRealNVP,favoni2020lattice,toth2020hamiltonian}, %
generative models
\citep{albergo2019flow,albergo2021introduction,medvidovic2020generative,boyda2020sampling,kanwar2020equivariant,wehenkel2020you,tomiya2021gauge},
and novel MCMC approaches
\citep{pasarica2010adaptively,tanaka2017towards,hoffman2019neutra,neklyudov2020orbital,neklyudov2020involutive,rezende2020normalizing,li2020neural}

\section{\label{sec:method}DLHMC Algorithm}
We provide a review of the generic Hamiltonian Monte Carlo (HMC) algorithm and set up some of the relevant notation in
\Secref{subsec:HMC}.
For simulating a system, $x$, using a theoretically given probability density $p(x)$ with likely intractable
normalization, we augment the Markov chain state as \(\xi = (x, v, d)\) with target distribution \(p(\xi) = p(x, v, d) =
p(x) p(v) p(d)\).
The conjugate momentum $v$ typically used in HMC algorithms has a known and easy to sample distribution.
The binary direction variable, as introduced in L2HMC~\citep{levy2017}, \(d\in\mathcal{U}(+,-)\), denoting the
``direction'' (forward/backward) of our update.

\begin{figure}[htpb]
   \centering
   \includegraphics[width=0.945\textwidth]{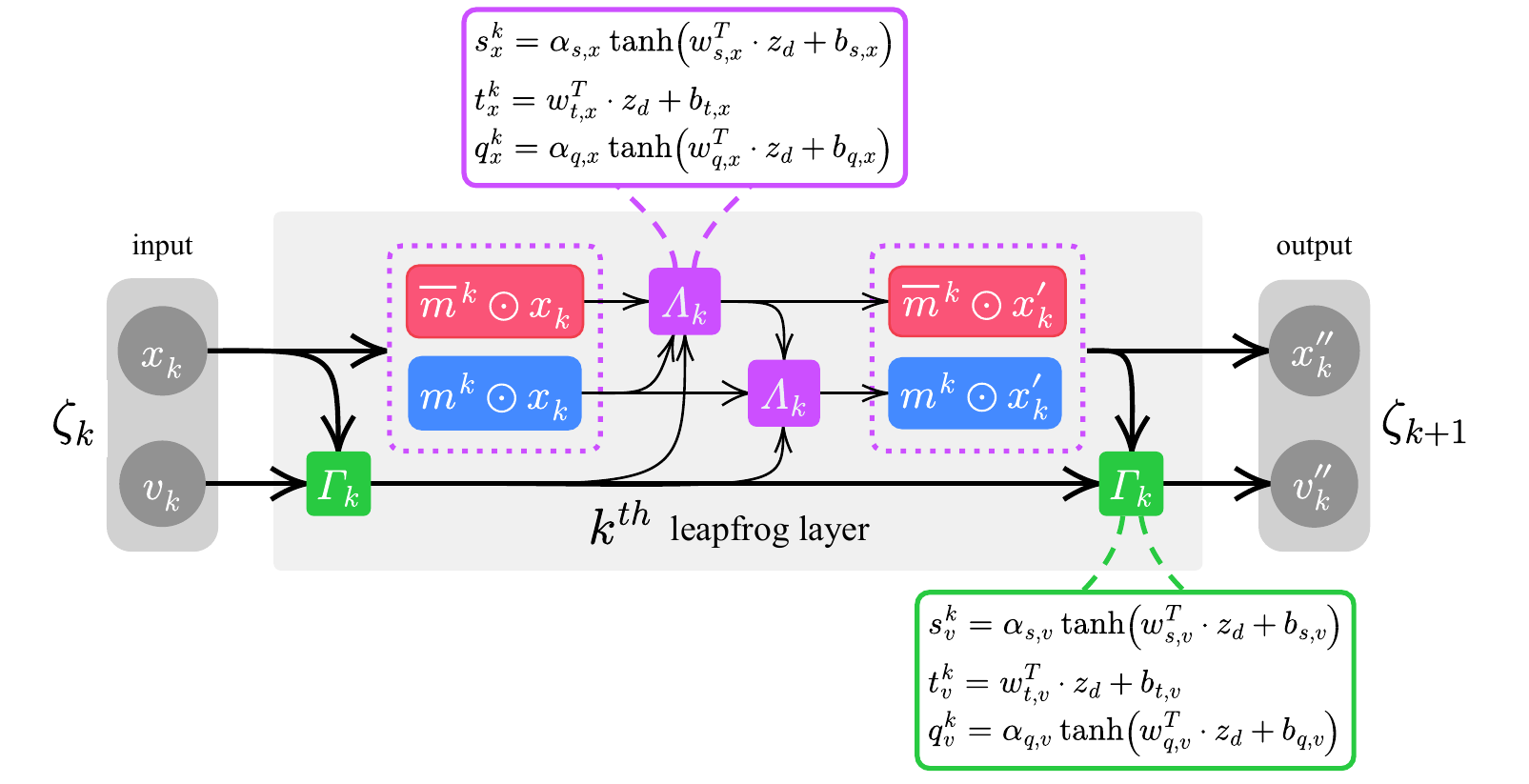}
   \caption{\label{fig:network}Illustration of the network architecture used in \Eqref{eq:newmomentumupdate} and
   \Eqref{eq:newpositionupdate}.}
\end{figure}

DLHMC further generalizes the leapfrog steps in L2HMC using individual neural networks, collectively parameterized by weights \(\theta\).
Each leapfrog step in DLHMC is a layer of the deep neural network that transforms the input $\xi_k=(x_k,v_k,d_k)$, for
the $k$-th layer, to output $(x''_k,v''_k,d_k)$, where $d_k$ stays constant choosing either the forward or backward
leapfrog layer.
Subsequently the $k+1$-th layer uses $\xi_{k+1}=(x''_k,v''_k,d_k)$ as its input.
For simplicity, we consider the forward \(d=+1\) direction, and introduce the notation:
\begin{align}
   v^{\prime}_{k} &\equiv \Gamma^{+}_{k}(v_{k};\zeta_{v_{k}})
   = v_{k}\odot \exp{\left(\tfrac{\varepsilon^{k}_{v}}{2}s_{v}^{k}(\zeta_{v_{k}})\right)} -
   \tfrac{\varepsilon^{k}_{v}}{2}{\left[\partial_{x}S(x_{k})\odot\exp{\left(\varepsilon^{k}_{v} q_{v}^{k}(\zeta_{v_{k}})\right)}
      +t_{v}^{k}(\zeta_{v_{k}})\right]},\label{eq:newmomentumupdate}\\
   x^{\prime}_{k} &\equiv \Lambda^{+}(x_{k};\zeta_{x_{k}})
   = x_{k}\odot\exp\left(\varepsilon^{k}_{x} s^{k}_{x}(\zeta_{x_{k}})\right)
   + \varepsilon^{k}_{x}\left[v^{\prime}_{k}\odot\exp\left(\varepsilon^{k}_{x} q^{k}_{x}(\zeta_{x_{k}})\right)
         + t^{k}_{x}(\zeta_{x_{k}})\right]\label{eq:newpositionupdate}
\end{align}
where: (1.) \(\zeta_{v_{k}} = (x_{k}, \partial_{x}S(x_{k}))\) and \(\zeta_{x_{k}} = (\mbart\odot x_{k}, v_{k})\),
\(\zeta_{x_{k}^{\prime}} = (\mt\odot x_{k}^{\prime}, v_{k}^{\prime})\) denote the (\(x\),
\(v\) respectively) networks' inputs\footnote{%
   The \(v\) network is independent of the variables being updated,
   the $x$ network will be made independent with the mask $m^k$.
},
where $\odot$ denotes element-wise multiplication; %
(2.) we indicate the direction of the update by the superscript \(\pm\) on \(\Gamma^{\pm}_{k}, \Lambda^{\pm}_{k}\); %
(3.) \(k\in\{0,1,\ldots,N_{\mathrm{LF}}-1\}\) denotes the current leapfrog step along the trajectory.
We include in \Figref{fig:network} an illustration of the network architecture of the $k$-th leapfrog layer used for the
updates in \Eqref{eq:newmomentumupdate} and \Eqref{eq:newpositionupdate}.

Using this notation, we can write a complete leapfrog update (in the forward \(d=+1\) direction)\footnote{%
   To obtain the expression for the reverse direction, we can invert each of the
   \(\Gamma^{-}\equiv{\left(\Gamma^{+}\right)}^{-1}, \Lambda^{-}\equiv{\left(\Lambda^{+}\right)}^{-1}\) functions, and
   perform the updates in the opposite order.
} as:
\begin{enumerate}
   \item Half-step momentum update:%
      \hspace{29pt}\(%
         v^{\prime}_{k} = \Gamma^{+}_{k}(v_{k};\zeta_{v_{k}})%
   \)
   \item Full-step half-position update:\footnote{%
         By this we mean we are performing a complete update on one-half of the indices of \(x\) (determined by
         \(\mt\odot x\)), followed by an analogous update of the complementary indices, \(\mbart\odot x\).
   }
      \hspace{14pt} \(%
         x^{\prime}_{k} = \mbart\odot x_{k} + \mt\odot \Lambda^{+}_{k}(x_{k};\zeta_{x_{k}})
   \)
   \item Full-step half-position update:%
      \hspace{21pt} \(%
         x^{\prime\prime}_{k} = \mbart\odot\Lambda^{+}_{k}(x^{\prime}_{k};\zeta_{x^{\prime}_{k}}) + \mt\odot x^{\prime}_{k}
   \)
   \item Half-step momentum update:%
      \hspace{25pt} \(%
         v^{\prime\prime}_{k} = \Gamma_{k}^{+}(v^{\prime}_{k}; \zeta_{v^{\prime}_{k}})
   \)
\end{enumerate}
Collectively, we refer to this series of updates as a single \emph{leapfrog layer}, which performs a single update
\(\xi\rightarrow\xi^{\prime}\).
Note that in order to keep our update reversible, we've split the \(x\) update into two sub-updates by introducing a
binary mask \(\mbart = \mathbbm{1} - \mt\) that updates half of the components of \(x\) sequentially.

As in HMC, we form a complete trajectory by performing \(N_{\mathrm{LF}}\) leapfrog steps in sequence, followed by a
Metropolis-Hastings accept/reject step as described in \Eqref{eq:mhcriteria}.
However, unlike in the expression for HMC, we must take into account the overall Jacobian factor from the update
\(\xi\rightarrow\xi^{\prime}\), which can be easily computed as \(\left|\tfrac{\partial v^{\prime\prime}_{k}}{\partial
   v_{k}}\right| = \exp{\left(\tfrac{1}{2}{\varepsilon^{k}_{v} s^{k}_{v}(\zeta_{v_{k}})}\right)}\),
   \(\left|\tfrac{\partial x^{\prime\prime}_{k}}{\partial x_{k}}\right| = \exp{\left(\varepsilon^{k}_{x} s^{k}_{x}(\zeta_{x_{k}})\right)}\).
where we keep the Jacobian factor for individual degree of freedom tractable.
\section{\label{sec:latticegaugetheory}\(U(1)\) Lattice Gauge Theory}
\begin{wrapfigure}{r}{.25\textwidth}
      \centering
      \includegraphics[width=0.25\textwidth]{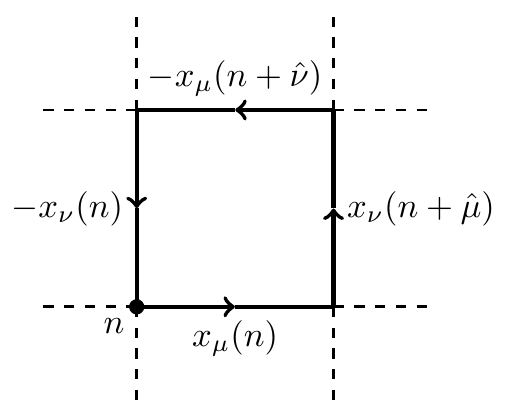}
      \caption{\label{fig:plaquette}Plaquette}
\end{wrapfigure}
Lattice gauge theory is one formulation of quantum field theories with gauge fields, eg. Quantum Electrodynamics (QED)
and Quantum Chromodynamics (QCD).
It is the predominant method that can be computed numerically, and widely used for simulating subatomic particles and
nuclei.
Here we describe the \(U(1)\) gauge (QED), and apply our DLHMC in 2 dimensional systems.
Let \(U_{\mu}(n) = e^{i x_{\mu}(n)} \in U(1)\), with \(x_{\mu}(n) \in [-\pi,\pi]\) denote the \emph{link variables},
where \(x_{\mu}(n)\) is the link oriented in the \(\hat{\mu}\)-direction located at the site \(n\).
We can write our target distribution in terms of the Wilson action \(S_{\beta}(x)\) as
\begin{equation}
   p_{t}(x) = \frac{1}{Z}e^{-\gamma_{t} S_{\beta}(x)},\quad\text{with}\quad S_{\beta}(x) = \beta \sum_{P}1 - \cos(x_{P})
   \label{eq:wilsonaction}
\end{equation}
where $Z$ is the intractable normalization;
\(\gamma_{t}\) is a scaling factor (\(\|\gamma_{t}\|\le 1\)) slowly annealed during training (\Secref{sec:annealing});
\(x_{P} \equiv x_{\mu}(n) + x_{\nu}(n+\hat{\mu}) - x_{\mu}(n+\hat{\nu}) -x_{\nu}(n)\) is a combination of link
variables around the \(1\times1\) elementary plaquette, shown in \Figref{fig:plaquette}; and the sum is over all
such plaquettes on the lattice.
Here, \(\beta = 2 / g_{0}^{2}\) is a coupling constant and \(\beta\rightarrow\infty\) recovers the continuum limit of
the theory. 

Physically, each lattice configuration has its topological charge, \(\mathcal{Q}_{\mathbb{Z}}\in\mathbb{Z}\), defined as
\begin{equation}
      \mathcal{Q}_{\mathbb{Z}}(x) \equiv \frac{1}{2\pi}\sum_{P}\left\lfloor x_{P}\right\rfloor,
   \quad\text{where}\quad \left\lfloor x_{P}\right\rfloor = x_{P} -
   2\pi\left\lfloor\frac{x_{P}+\pi}{2\pi}\right\rfloor.
   \label{eq:intcharge}
\end{equation}
Current methods are severely limited in their ability to mix between different topologies, a phenomenon known as
topological freezing, when simulation approaches the continuum limit.
For this reason, we wish to construct a loss function that encourages our sampler to explore different topological sectors.
In order to do so, we introduce a continuous version of the topological charge,
\(\mathcal{Q}_{\mathbb{R}}\in\mathbb{R}\), by replacing the projection in \Eqref{eq:intcharge} to give:
\begin{equation}
    \mathcal{Q}_{\mathbb{R}} \equiv \frac{1}{2\pi}\sum_{P}\sin(x_{P}).
    \label{eq:sincharge}
\end{equation}
This quantity has the advantage of being continuously differentiable, which is important for training the deep neural
network.
Our loss function is then defined as
\begin{equation}
   \mathcal{L}(\theta) = \mathbb{E}_{p_{t}(\xi)}{%
      \left[\,-\delta(\xi^{\prime}, \xi) \, A(\xi^{\prime}|\xi)\,\right]
   }
\end{equation}
where \(\delta(\xi^{\prime}, \xi) \equiv {\left(\mathcal{Q}_{\mathbb{R}}(x^{\prime}) -
\mathcal{Q}_{\mathbb{R}}(x)\right)}^{2}\) and \(A(\xi^{\prime}|\xi)\) is given in \Eqref{eq:mhcriteria}.
In practice we approximate the exact loss function with a sample
over a batch of MC configurations.
\section{\label{sec:results}Results}
We apply our DLHMC algorithm with
the training procedure in \Secref{sec:algorithm} and the annealing schedule from \Secref{sec:annealing}
to the 2D $U(1)$ gauge theory with parameters gradually approaching the continuum limit.
We then run inference to generate the Markov chain using the trained model and compare its efficiency with HMC.\@
We trained our models using Horovod \citep{horovod2018sergeev} with TensorFlow \citep{tensorflow2015-whitepaper} on the
ThetaGPU supercomputer at the Argonne Leadership Computing Facility.
A typical training run on 1 node (\(8\times\) NVIDIA A100 GPUs) using a batch size \(M=2048\), hidden layer shapes
\(=\left[256, 256, 256\right]\) for each of the \(N_{\mathrm{LF}}=10\) leapfrog layers, on a \(16\times16\) lattice, for
\(5\times10^{5}\) training steps takes roughly \(24\) hours to complete.

\begin{wrapfigure}{r}{0.425\textwidth}
   \centering
      \centering
      \includegraphics[width=0.425\textwidth]{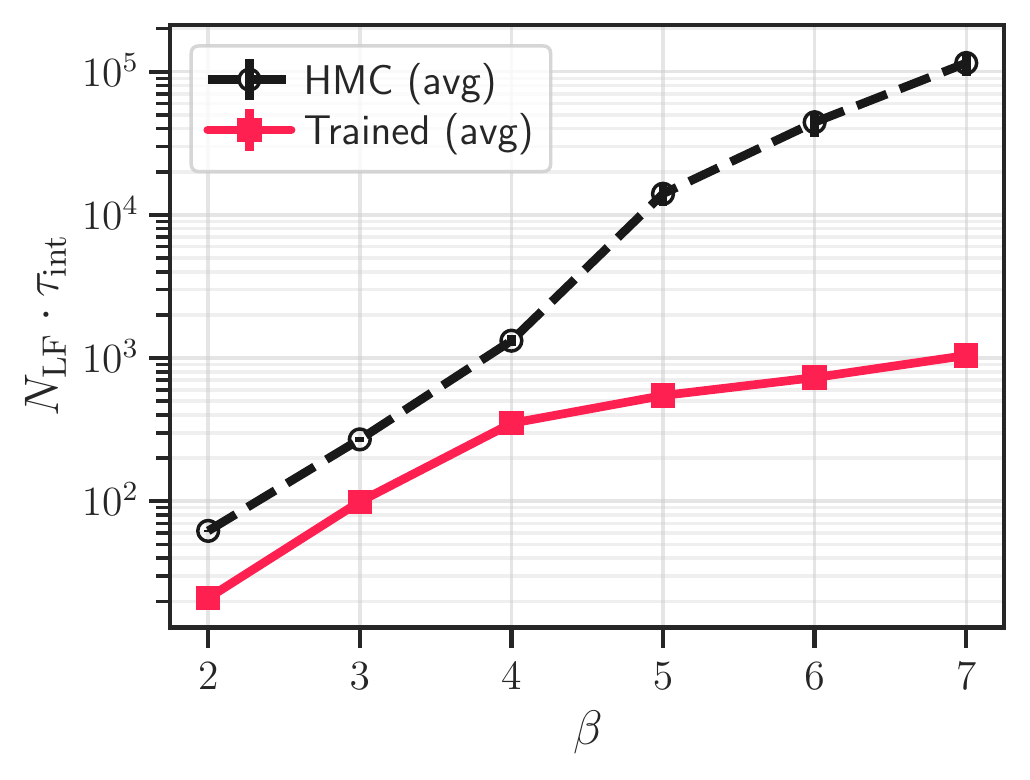}
      \caption{\label{fig:autocorrvsbeta}Estimate of the integrated autocorrelation time
      \(\tau_{\mathrm{int}}^{\mathcal{Q}_{\mathbb{R}}}\) vs \(\beta\), scaled by \(N_{\mathrm{LF}}\) to account for
   simulation cost.}
\end{wrapfigure}
\begin{figure}[htpb]
   \centering
   \includegraphics[width=\textwidth]{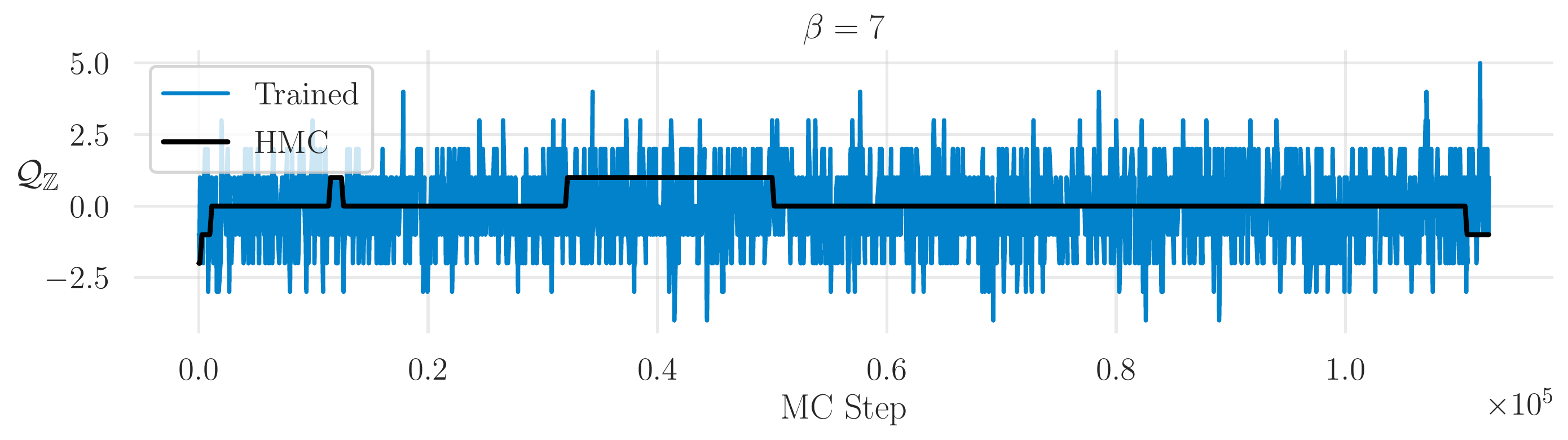}
   \caption{\label{fig:topfreeze}Topological charge \(\mathcal{Q}_{\mathbb{Z}}\) vs MC
   step for both HMC (black) and the trained model (blue).}
\end{figure}
\begin{figure}[hbpt]
   \begin{subfigure}[t]{0.33\textwidth}
      \includegraphics[width=\textwidth]{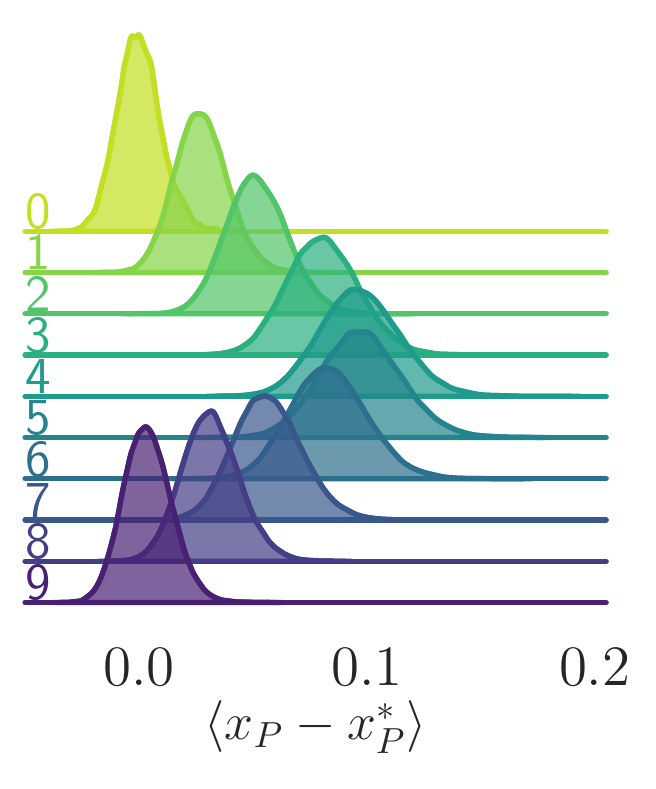}
      \caption{\label{fig:plaqsf}Difference in the average plaquette, \(\langle x_{P}-x_{P}^{*}\rangle\) at each
      leapfrog layer.}
   \end{subfigure}
   \hfill
   \begin{subfigure}[t]{0.315\textwidth}
      \includegraphics[width=\textwidth]{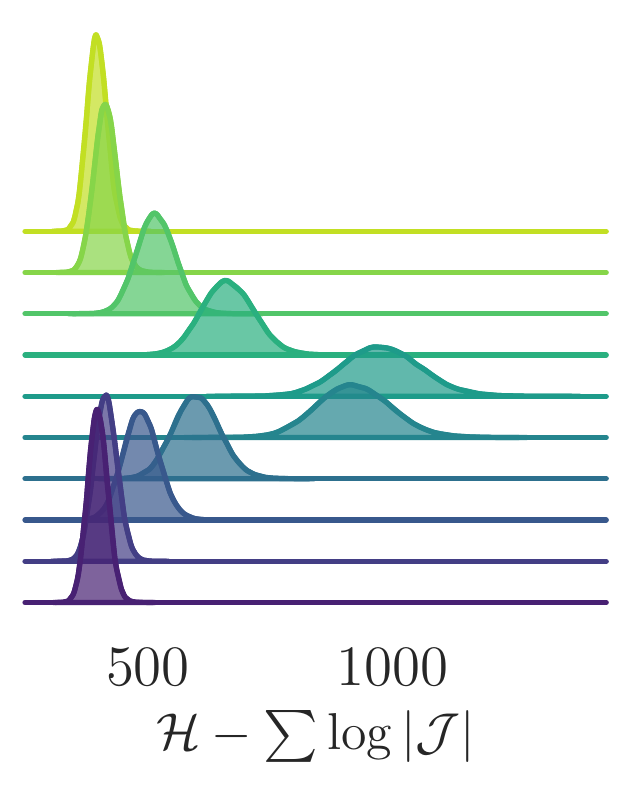}
      \caption{\label{fig:hwf}The adjusted energy, \(\mathcal{H}-\sum\log|\mathcal{J}|\) at each leapfrog layer.}
   \end{subfigure}
   \hfill
   \begin{subfigure}[t]{0.315\textwidth}
      \includegraphics[width=\textwidth]{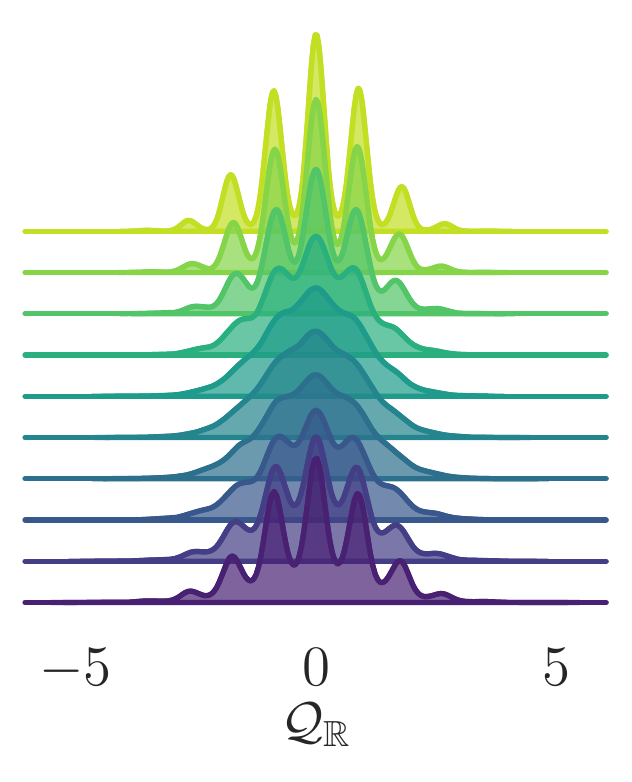}
      \caption{\label{fig:sinQf}The continuous topological charge, \(\mathcal{Q}_{\mathbb{R}}\) at each
      leapfrog layer.}%
   \end{subfigure}
   \caption{Plots showing how various quantities of interest are transformed as they're passed through each of the
   leapfrog layers in the generalized leapfrog update.}
\end{figure}

To measure the improvement of our model, we evaluate the integrated autocorrelation time of the topological charge, $\tau_{\mathrm{int}}^{\mathcal{Q}_{\mathbb{R}}}$, which
can be interpreted roughly as the number of trajectories needed (on average) before an independent sample is drawn.
In order to more accurately capture the computational effort between the two approaches we scale our estimate of the
integrated autocorrelation time by the number of leapfrog steps as \(N_{\mathrm{LF}}\tau_{\mathrm{int}}^{\mathcal{Q}_{\mathbb{R}}}\).
We can see in
\Figref{fig:autocorrvsbeta}, that the trained model consistently outperforms generic HMC across \(\beta = 2, 3, \ldots, 7\) and for the largest
lattice coupling studied, $\beta=7$, standard HMC would take about one hundred times more updates to generate an independent configuration as DLHMC.
To ensure the validity of our results, each trained run was compared to multiple HMC runs across a range of
\(N_{\mathrm{LF}}\) and \(\varepsilon\) values.
We see in \Figref{fig:topfreeze} that HMC remains stuck at a particular value of \(\mathcal{Q}_{\mathbb{Z}}\) for large
sections of the simulation, whereas the trained sampler rapidly jumps between values.
In order to better understand the mechanism driving this improved behavior, we evaluate different quantities of the
system during its passing through the $N_{\mathrm{LF}}$ leapfrog layers.
\Figref{fig:plaqsf} and \Figref{fig:hwf} show the distribution over batches of the average plaquette value and the
modified energy of the Markov state, during the transformation via $N_{\mathrm{LF}}=10$ leapfrog layers.
Our sampler artificially increased the energy density of the physical system during the first half of the trajectory
before returning back to its original physical value.  This is not a prescribed behavior, but self-learned during the
training.
We believe that this ability to vary the energy during the trajectory helps the sampler to overcome energy barriers
between topological sectors whereas HMC remains stuck.
As can be seen in \Figref{fig:sinQf}, the continuous topological charge exhibits mixing behavior at the middle of the
deep neural network.
\section{Discussion and Further Research}
We propose DLHMC as an efficient algorithm for MCMC, and observed a dramatic improvement in simulating 2D $U(1)$ gauge
theory.
Being able to efficiently tunnel topological sectors is a significant first step toward efficient simulating lattice
gauge theories describing our universe.
We plan to continue developing this approach with the goal of eventually scaling up to the 4D \(SU(3)\) gauge theory of
QCD.\@
\subsubsection*{Acknowledgments}
This research was supported by the Exascale Computing Project (17-SC-20-SC), a collaborative effort of the U.S.
Department of Energy Office of Science and the National Nuclear Security Administration.
This research was performed using resources of the Argonne Leadership Computing Facility (ALCF), which is a DOE Office
of Science User Facility supported under Contract DE\_AC02--06CH11357. 
This work describes objective technical results and analysis.
Any subjective views or opinions that might be expressed in the work do not necessarily represent the views of the U.S.
DOE or the United States Government.
Results presented in this research were obtained using the Python \citep{van1995python}, programming language and its
many data science libraries
\citep{matplotlib,numpyharris2020array,tensorflow2015-whitepaper,seaborn_michael_waskom_2017_883859,ipython4160251,arviz_2019}

\bibliography{iclr2021_conference.bib}
\bibliographystyle{iclr2021_conference}

\appendix
\section{Appendix}
\subsection{\label{subsec:HMC}Hamiltonian Monte Carlo (HMC)}
The Hamiltonian Monte Carlo algorithm is a widely used technique that allows us to sample from an analytically known
target distribution \(p(x)\) by constructing a chain of states \(\{x^{(0)}, x^{(1)}, \ldots, x^{(n)}\}\), such that
\(x^{(n)}\sim p(x)\) in the limit \(n\rightarrow\infty\).
For our purposes, we assume that our target distribution can be expressed as a Boltzmann distribution, \(p(x) =
\tfrac{1}{\mathcal{Z}} e^{-S(x)}\propto e^{-S(x)}\), where \(S(x)\) is the \emph{action} of our theory, and
\(\mathcal{Z}\) is a normalization factor (the partition function).
In this case, HMC begins by augmenting the state space with a fictitious momentum variable \(v\), normally
distributed independently of \(x\), i.e.\ \(v\sim\mathcal{N}(0, \mathbbm{1})\).
Our joint distribution can then be written as \(%
   p(x, v) = p(x) p(v) \propto e^{-S(x)} e^{-\frac{1}{2}v^{T}v} = e^{-\mathcal{H}(x, v)}
\), where \(\mathcal{H}(x, v)\) is the Hamiltonian of the joint (x, v) system.
This system obeys Hamilton's equations: %
\(\dot{x} = \frac{\partial\mathcal{H}}{\partial v}\), \(\dot{v} = -\frac{\partial H}{\partial x}\), which can be 
numerically integrated along iso-probability contours of \(\mathcal{H} = \text{const.}\).
Explicitly, for a step size \(\varepsilon\) and initial state \(\xi = (x, v)\), the leapfrog integrator generates a
proposal configuration \(\xi^{\prime} \equiv (x^{\prime}, v^{\prime})\) by performing the following series of updates: 
%
%
\begin{enumerate}
   \item Half-step momentum update: \hspace{12pt}\(%
      v^{1/2} \equiv v{\left(t+\frac{\varepsilon}{2}\right)} = v-\frac{\varepsilon}{2}\partial_{x}S(x)
   \)
   \item Full-step position update: \hspace{36pt}\(%
      x^{\prime} \equiv x(t+\varepsilon) = x + \varepsilon v^{1/2}
   \)
   \item Half-step momentum update:
      \hspace{18pt} \(%
         v^{\prime} \equiv v(t+\varepsilon) = v^{1/2} - \frac{\varepsilon}{2}\partial_{x} S(x^{\prime})
   \)
\end{enumerate}
We can then construct a complete \emph{trajectory} of length \(\lambda = \varepsilon N_{\mathrm{LF}}\) by
performing \(N_{\mathrm{LF}}\) leapfrog steps in sequence.
At the end of our trajectory, we either accept or reject the proposal configuration according to the Metropolis-Hastings
acceptance criteria,
\begin{equation}
   x_{i+1} =
   \begin{cases}%
      x^{\prime} &\mbox{with probability } A(\xi^{\prime}|\xi) \\
      x &\mbox{with probability } (1 - A(\xi^{\prime}|\xi)), \quad\text{and}\quad%
         A(\xi^{\prime}|\xi) = \min\left\{%
            1, \frac{p(\xi^{\prime})}{p(\xi)}\left|\frac{\partial{\xi^{\prime}}}{\partial\xi^{T}}\right|%
         \right\}.
   \end{cases}
   \label{eq:mhcriteria}
\end{equation}
The leapfrog integrator is symplectic, so the Jacobian factor reduces to
\(\left|\frac{\partial\xi^{\prime}}{\partial\xi^{T}}\right| = 1\). 
\section{\label{sec:annealing}Annealing Schedule}
To help our sampler overcome the large energy barriers between isolated modes, we introduce an \emph{annealing schedule}
during the training phase (\(N\) training steps) \({\{\gamma_{t}\}}_{t=0}^{N} = \{\gamma_{0}, \gamma_{1}, \ldots,
\gamma_{N-1}, \gamma_{N}\}\), where \(\gamma_{0} < \gamma_{1} < \cdots < \gamma_{N} \equiv 1\), \(\gamma_{t+1} -
\gamma_{t} \ll 1\).
We are free to vary \(\gamma\) during the initial training phase as long as we recover the true distribution
with \(\gamma \equiv 1\) at the end of training and evaluate our trained model without this factor.
Explicitly, for \(\gamma_{t} < 1\) this rescaling factor helps to reduce the height of the energy barriers, making it
easier for our sampler to explore previously inaccessible regions of the phase space.
In terms of this additional annealing schedule, our target distribution picks up an additional index \(t\) to represent
our progress through the training phase, which can be written explicitly as 
\begin{equation}
   p_{t}(x)\propto e^{-\gamma_{t} S(x)},\quad\text{for}\quad t=0, 1, \ldots, N
   \label{eq:targetannealing}
\end{equation}
\section{\label{sec:algorithm}Training Algorithm}
\begin{algorithm}[htpb]%
   \SetAlgoLined%
   \SetAlgoVlined%
   \SetKwProg{Fn}{def}{\string:}{}%
   \SetKwFunction{Range}{range}%
   \SetKwFor{For}{\color{newblue}\textbf{\texttt{for}}}{\string:}{}\color{black}%
   \SetKwIF{If}{ElseIf}{Else}{if}{:}{elif}{else:}{}%
   \SetKwFor{While}{while}{:}{fintq}%
   \SetKwInOut{Input}{\color{pinkred}{\textbf{\texttt{input}}}\color{black}}%
   \AlgoDontDisplayBlockMarkers\SetAlgoNoEnd%
   \DontPrintSemicolon%
   \caption{\label{alg:training}Training procedure}%
   \Input{%
      \begin{enumerate}
         \item \texttt{Loss function}, \(\mathcal{L}_{\theta}(\xi^{\prime},\xi,
            A(\xi^{\prime}|\xi))\)
         \item \texttt{Batch of initial states}, \(x\)
         \item \texttt{Learning rate schedule}, \({\{\alpha_{t}\}}_{t=0}^{N_{\mathrm{train}}}\)
         \item \texttt{Annealing schedule}, \({\{\gamma_{t}\}}_{t=0}^{N_{\mathrm{train}}}\)
         \item \texttt{Target distribution,} \(p_{t}(x)\propto e^{-\gamma_{t} S_{\beta}(x)}\)
      \end{enumerate}
   }%
   \texttt{Initialize weights} \(\theta\)\;
   \For{\(0 \leq t < N_{\mathrm{train}}\)}{%
      \texttt{resample } \(v\sim\mathcal{N}(0, \mathbbm{1})\)\;
      \texttt{resample } \(d\sim\mathcal{U}(+,-)\)\;
      \texttt{construct } \(\xi_{0} \equiv (x_{0}, v_{0}, d_{0})\)\;
      \For{\(0 \leq k < N_{\mathrm{LF}}\)}{%
         \texttt{propose (leapfrog layer) } \(\xi_{k}^{\prime}\leftarrow \xi_{k}\)
      }
      \texttt{compute } \(A(\xi^{\prime}|\xi) =%
      \min\left\{1,\frac{p(\xi^{\prime})}{p(\xi)}\left|\frac{\partial%
      \xi^{\prime}}{\partial \xi^{T}}\right|\right\}\)\;
      \texttt{update } \(\mathcal{L}\leftarrow \mathcal{L}_{\theta}(\xi^{\prime},\xi,%
      A(\xi^{\prime}|\xi))\)\;
      \texttt{backprop } \(\theta\ \leftarrow\ \theta-\alpha_t \nabla_{\theta} \mathcal{L}\)\;
      \texttt{assign } \(x_{t+1} \leftarrow
      \begin{cases}
         x^{\prime} &\mbox{with probability } A(\xi^{\prime}|\xi) \\
         x &\mbox{with probability } (1 - A(\xi^{\prime}|\xi)).%
      \end{cases}%
      \)\;
   }\;
\end{algorithm}%
%
%
\end{document}

%% file: math_commands.tex
\usepackage{amsmath,amsfonts,bm}

\def\Figref#1{Figure~\ref{#1}}

\def\Secref#1{Section~\ref{#1}}

\def\eqref#1{equation~\ref{#1}}
\def\Eqref#1{Equation~\ref{#1}}

\def\1{\bm{1}}

\DeclareMathAlphabet{\mathsfit}{\encodingdefault}{\sfdefault}{m}{sl}
\SetMathAlphabet{\mathsfit}{bold}{\encodingdefault}{\sfdefault}{bx}{n}

